\documentclass[10pt,twocolumn,letterpaper]{article}
\usepackage{marvosym}

\usepackage{cvpr}              % To produce the CAMERA-READY version
\usepackage[dvipsnames]{xcolor}

\definecolor{cvprblue}{rgb}{0.21,0.49,0.74}
\usepackage[pagebackref,breaklinks,colorlinks,citecolor=cvprblue]{hyperref}

\def\paperID{*****} % *** Enter the Paper ID here
\def\confName{CVPR}
\def\confYear{2024}

\title{Domain Adaptation Using Pseudo Labels for COVID-19 Detection}

\author{Runtian Yuan$^1$, Qingqiu Li$^1$, Junlin Hou$^2$, Jilan Xu$^1$, \\
Yuejie Zhang$^1$, Rui Feng$^1$, Hao Chen$^2$\\
\\
$^1$ Fudan University\\
$^2$ The Hong Kong University of Science and Technology\\
}

\begin{document}
\maketitle
\begin{abstract}
In response to the need for rapid and accurate COVID-19 diagnosis during the global pandemic, we present a two-stage framework that leverages pseudo labels for domain adaptation to enhance the detection of COVID-19 from CT scans. By utilizing annotated data from one domain and non-annotated data from another, the model overcomes the challenge of data scarcity and variability, common in emergent health crises. The innovative approach of generating pseudo labels enables the model to iteratively refine its learning process, thereby improving its accuracy and adaptability across different hospitals and medical centres. Experimental results on COV19-CT-DB database showcase the model's potential to achieve high diagnostic precision, significantly contributing to efficient patient management and alleviating the strain on healthcare systems. Our method achieves 0.92 Macro F1 Score on the validation set of Covid-19 domain adaptation challenge.
\end{abstract}    
\section{Introduction}
\label{sec:intro}

% The outbreak of the novel coronavirus, officially named COVID-19 (Coronavirus Disease 2019) by the World Health Organization, has posed an unprecedented global health crisis. The causative agent of COVID-19, SARS-CoV-2 (Severe Acute Respiratory Syndrome Coronavirus 2), is a member of the coronavirus family, which includes viruses responsible for common colds as well as more severe illnesses such as SARS (Severe Acute Respiratory Syndrome) and MERS (Middle East Respiratory Syndrome). 
The COVID-19 (Coronavirus Disease 2019) pandemic, caused by the novel coronavirus SARS-CoV-2, has necessitated rapid and accurate diagnostic methods to manage and control its spread effectively.
% In light of the urgent need for rapid and accurate diagnostic methods, Computed Tomography (CT) imaging has emerged as a critical diagnostic tool in the battle against COVID-19. CT scans provide a detailed view of the lungs and can detect characteristic signs of the infection, such as ground-glass opacities and bilateral pulmonary infiltrates, even in the absence of symptoms or before the detection of the virus by RT-PCR tests. However, the reliance on manual analysis of CT images presents challenges, including the need for specialized expertise and the risk of human error, especially under high-pressure conditions.
Given the vast and rapid spread of COVID-19, efficient and accurate diagnostic methods are crucial for controlling the outbreak. Computed Tomography (CT) imaging has proven to be an invaluable tool in the early detection and management of the disease, capable of revealing pulmonary manifestations such as ground-glass opacities and bilateral infiltrates characteristic of COVID-19. However, the interpretation of CT images requires significant expertise and can be time-consuming, presenting challenges in high-demand scenarios and potentially leading to delays in diagnosis and treatment.

Recent deep-learning methods have been widely explored in solving COVID-19 detection~\cite{kollias2020deep,kollias2021mia,kollias2023deep,hou2021periphery,subramanian2022review,hou2022cmc_v2,hou2022boosting}. However, most of them require large amount of annotated data, which are hard to acquire in real-world scenarios.
Confronted with the challenge of scarce annotated data, we employ pseudo labels for domain adaptation to enhance COVID-19 detection from CT images. Our approach leverages annotated data from one domain (Domain A) and harnesses the unannotated or sparsely annotated data from another domain (Domain B), addressing the common challenge of data scarcity in emergent health crises. By generating pseudo labels for the unlabeled data in Domain B, the model iteratively refines its learning, bridging the gap between the two domains and improving its predictive accuracy and robustness.
The use of pseudo labels in domain adaptation represents a strategic method to mitigate the label scarcity problem by assigning provisional labels to unlabeled data, thereby expanding the training dataset. This technique not only facilitates the transfer of knowledge from the source domain with abundant annotated data to the target domain with limited or no annotations but also enhances the model's generalizability across diverse clinical settings and populations.

In this paper, we present our framework for the second challenge of the 4th Covid-19 competition~\cite{kollias2024domain}. The proposed model aims to achieve a high level of diagnostic precision while significantly reducing the time and resources required for COVID-19 detection via CT scans. This advancement holds the promise of enabling faster, more accurate, and scalable diagnostic solutions.

\section{Methodology}
\label{sec:method}

\begin{figure*}
    \centering
    \includegraphics[width=1\linewidth]{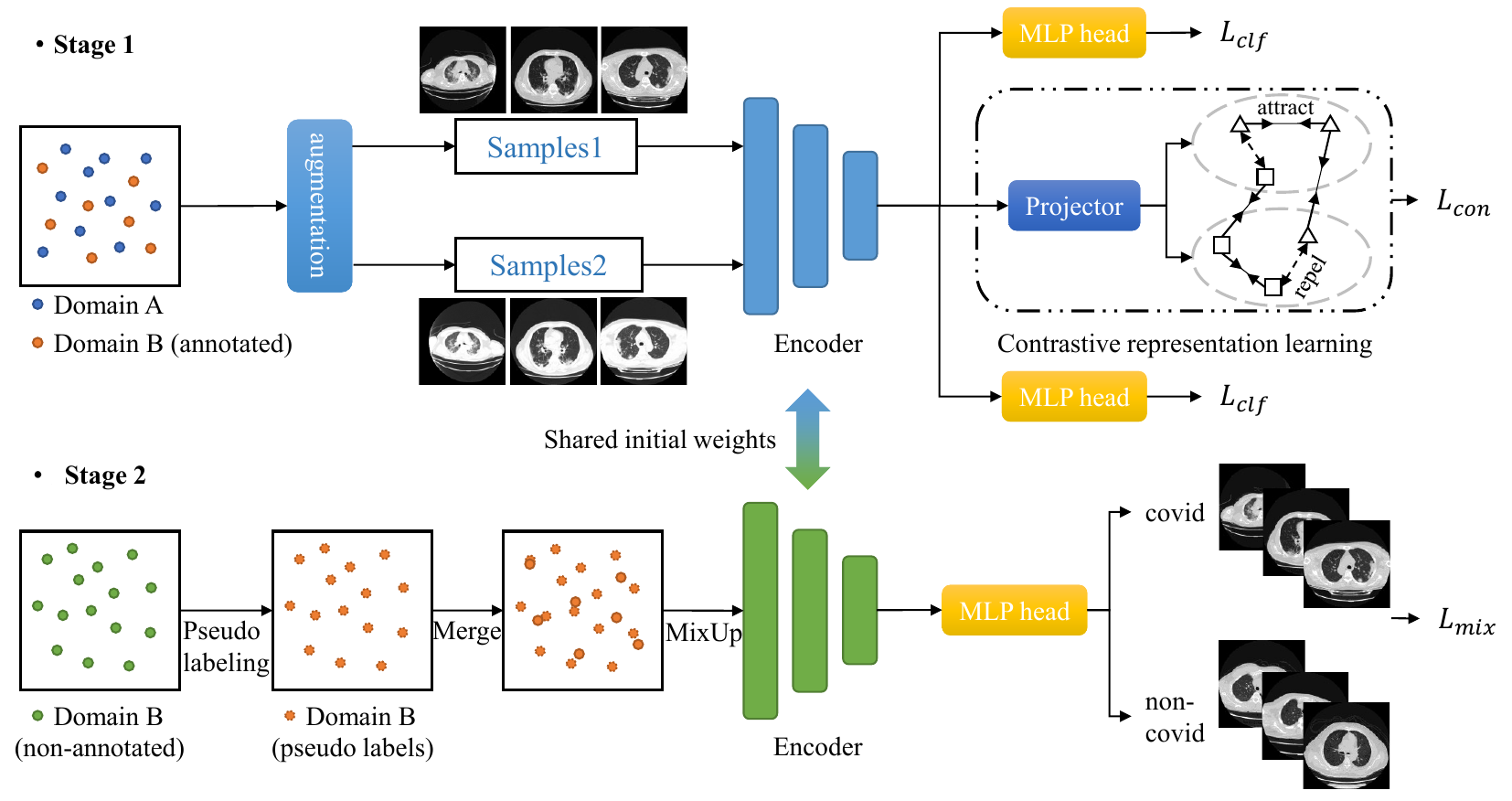}
    \caption{Overview of our framework for COVID-19 domain adaptation.}
    \label{fig:arch}
\end{figure*}

% Fig.\ref{fig:arch} illustrates the overview of our framework. The framework is divided into two stages. The first stage is to train a model on annotated data from both domain A and B using augmentation and contrastive representation learning. The second stage is to take the model in stage 1 to geneate pseudo labels for non-annotated data from domain B, and then merge with annotated data to detect COVID-19. We adopt our previous work, the CMC network~\cite{hou2021cmc}, as the baseline method, which has shown excellent COVID-19 detection performance.
Figure \ref{fig:arch} provides a schematic representation of our framework, which is structured into two stages. In the initial stage, the model is trained on annotated data from both Domain A and Domain B, employing techniques such as data augmentation and contrastive representation learning to enhance its learning efficacy. The subsequent stage involves leveraging the pre-trained model from the first stage to generate pseudo labels for the non-annotated data in Domain B. This augmented dataset, comprising both pseudo-labeled and originally annotated data, is then utilized for improved COVID-19 detection. Our methodology builds upon the foundation laid by our previous work with the CMC network~\cite{hou2021cmc}, which has already demonstrated its excellent performance in accurately detecting COVID-19, serving as a robust baseline for the current framework.

\section{Datasets}

COV19-CT-DB~\cite{kollias2023deep} contains 3D chest CT scans that collected in various medical centers. The database includes 7,756 3-D CT scans; 1,661 are COVID-19 samples, whilst 6,095 refer to non COVID-19 ones. There are about 2,500,000 images included in these datasets. All have been anonymized. 724,273 images refer to the COVID-19 class, whilst 1,775,727 slices belong to non COVID-19 class~\cite{arsenos2022large,arsenos2023data,kollias2020deep,kollias2020transparent,kollias2021mia,kollias2022ai,kollias2023ai}.
For the COVID-19 domain adaptation challenge, the training set contains 239 annotated 3D CT scans (120 COVID-19 cases and 119 Non-COVID-19 cases) and 494 non-annotated 3D CT scans. The validation set consists of 178 3D CT scans (65 COVID-19 cases and 113 Non-COVID-19 cases). 
The testing set includes 4055 scans and the labels are not available during the challenge.
\section{Experiments}
\subsection{Pre-Processing}
Our data pre-processing procedure is as follows. First, each 2D chest CT scan series is composed into a 3D volume of shape (D, H, W), where D, H, W denotes the slice, height, and width, respectively. Then, each volume is resized from its original size to (128, 256, 256). Finally, we transform the CT volume to the interval [0, 1] for intensity normalization. 

\subsection{Implementation Details}
We apply inflated 3D ResNest50 as the backbones in our experiments. The value of parameter $d_e$ is 2,048 and $d_p$ is 128. 
We optimize the network using the Adam algorithm with a weight decay of $10^{-5}$. The network is trained for 100 epochs. The initial learning rate is set to 0.0001 and then divided by 10 at 30\% and 80\% of the total number of training epochs. Our methods are implemented in PyTorch and run on four NVIDIA Tesla V100 GPUs.
We adopt the Macro F1 Score as the evaluation metrics. The Macro F1 Score is defined as the unweighted average of the classwise/label-wise F1 Scores, i.e., the unweighted average of the COVID-19/non-COVID-19 class F1 Score.

\subsection{Experimental Results}
Table \ref{tab:val} shows the results of the baseline model and our method on the validation set of COVID-19 domain adaptation challenge.
The baseline model~\cite{kollias2024domain} employs Monte Carlo Dropout to assess uncertainty while training the CNN-RNN architecture using data from both A (annotated) and B (annotated).
We adopt the CMC architecture with Contrastive learning and Mixup, achieving the detection performance with 0.92 Macro F1 Score.
\begin{table}
    \centering
    \begin{tabular}{|c|c|}
        \hline
        Method & `macro’ F1 Score\\
        \hline
        Baseline~\cite{kollias2024domain} & 0.73 \\
        \hline
        \textbf{Ours} & \textbf{0.92} \\
        \hline
    \end{tabular}
    \caption{The comparison results on the validation set of Covid-19 domain adaptation challenge.}
    \label{tab:val}
\end{table}
\section{Conclusion}
In this paper, we propose a two-stage framework to navigate the challenges posed by the scarcity of annotated data in the realm of COVID-19 detection using CT scans. By employing pseudo labels for domain adaptation, our model not only capitalizes on the available annotated data from one domain but also harnesses the untapped potential of non-annotated data from another domain, significantly enhancing its diagnostic capabilities. The two-stage approach, beginning with training on augmented datasets followed by pseudo label generation for domain adaptation, demonstrates a marked improvement in detection accuracy. This advancement, rooted in the principles of our previously developed CMC network, underscores the potential of our framework to revolutionize COVID-19 diagnostics, offering a scalable and efficient solution that could substantially alleviate the pressures on healthcare systems worldwide.
{
    \small
    \bibliographystyle{ieeenat_fullname}
    \bibliography{main}

\begin{thebibliography}{14}
\providecommand{\natexlab}[1]{#1}
\providecommand{\url}[1]{\texttt{#1}}
\expandafter\ifx\csname urlstyle\endcsname\relax
  \providecommand{\doi}[1]{doi: #1}\else
  \providecommand{\doi}{doi: \begingroup \urlstyle{rm}\Url}\fi

\bibitem[Arsenos et~al.(2022)Arsenos, Kollias, and Kollias]{arsenos2022large}
Anastasios Arsenos, Dimitrios Kollias, and Stefanos Kollias.
\newblock A large imaging database and novel deep neural architecture for covid-19 diagnosis.
\newblock In \emph{2022 IEEE 14th Image, Video, and Multidimensional Signal Processing Workshop (IVMSP)}, page 1–5. IEEE, 2022.

\bibitem[Arsenos et~al.(2023)Arsenos, Davidhi, Kollias, Prassopoulos, and Kollias]{arsenos2023data}
Anastasios Arsenos, Andjoli Davidhi, Dimitrios Kollias, Panos Prassopoulos, and Stefanos Kollias.
\newblock Data-driven covid-19 detection through medical imaging.
\newblock In \emph{2023 IEEE International Conference on Acoustics, Speech, and Signal Processing Workshops (ICASSPW)}, page 1–5. IEEE, 2023.

\bibitem[Hou et~al.(2021{\natexlab{a}})Hou, Xu, Feng, Zhang, Shan, and Shi]{hou2021cmc}
Junlin Hou, Jilan Xu, Rui Feng, Yuejie Zhang, Fei Shan, and Weiya Shi.
\newblock Cmc-cov19d: Contrastive mixup classification for covid-19 diagnosis.
\newblock In \emph{Proceedings of the IEEE/CVF International Conference on Computer Vision}, pages 454--461, 2021{\natexlab{a}}.

\bibitem[Hou et~al.(2021{\natexlab{b}})Hou, Xu, Jiang, Du, Feng, Zhang, Shan, and Xue]{hou2021periphery}
Junlin Hou, Jilan Xu, Longquan Jiang, Shanshan Du, Rui Feng, Yuejie Zhang, Fei Shan, and Xiangyang Xue.
\newblock Periphery-aware covid-19 diagnosis with contrastive representation enhancement.
\newblock \emph{Pattern Recognition}, 118:\penalty0 108005, 2021{\natexlab{b}}.

\bibitem[Hou et~al.(2022{\natexlab{a}})Hou, Xu, Zhang, Wang, Zhang, Zhang, and Feng]{hou2022cmc_v2}
Junlin Hou, Jilan Xu, Nan Zhang, Yi Wang, Yuejie Zhang, Xiaobo Zhang, and Rui Feng.
\newblock Cmc\_v2: Towards more accurate covid-19 detection with discriminative video priors.
\newblock In \emph{European Conference on Computer Vision}, pages 485--499. Springer, 2022{\natexlab{a}}.

\bibitem[Hou et~al.(2022{\natexlab{b}})Hou, Xu, Zhang, Zhang, Zhang, and Feng]{hou2022boosting}
Junlin Hou, Jilan Xu, Nan Zhang, Yuejie Zhang, Xiaobo Zhang, and Rui Feng.
\newblock Boosting covid-19 severity detection with infection-aware contrastive mixup classification.
\newblock In \emph{European Conference on Computer Vision}, pages 537--551. Springer, 2022{\natexlab{b}}.

\bibitem[Kollias et~al.(2020{\natexlab{a}})Kollias, Bouas, Vlaxos, Brillakis, Seferis, Kollia, Sukissian, Wingate, and Kollias]{kollias2020deep}
Dimitrios Kollias, N Bouas, Y Vlaxos, V Brillakis, M Seferis, Ilianna Kollia, Levon Sukissian, James Wingate, and S Kollias.
\newblock Deep transparent prediction through latent representation analysis.
\newblock \emph{arXiv preprint arXiv:2009.07044}, 2020{\natexlab{a}}.

\bibitem[Kollias et~al.(2020{\natexlab{b}})Kollias, Vlaxos, Seferis, Kollia, Sukissian, Wingate, and Kollias]{kollias2020transparent}
Dimitris Kollias, Y Vlaxos, M Seferis, Ilianna Kollia, Levon Sukissian, James Wingate, and Stefanos~D Kollias.
\newblock Transparent adaptation in deep medical image diagnosis.
\newblock In \emph{TAILOR}, page 251–267, 2020{\natexlab{b}}.

\bibitem[Kollias et~al.(2021)Kollias, Arsenos, Soukissian, and Kollias]{kollias2021mia}
Dimitrios Kollias, Anastasios Arsenos, Levon Soukissian, and Stefanos Kollias.
\newblock Mia-cov19d: Covid-19 detection through 3-d chest ct image analysis.
\newblock In \emph{Proceedings of the IEEE/CVF International Conference on Computer Vision}, page 537–544, 2021.

\bibitem[Kollias et~al.(2022)Kollias, Arsenos, and Kollias]{kollias2022ai}
Dimitrios Kollias, Anastasios Arsenos, and Stefanos Kollias.
\newblock Ai-mia: Covid-19 detection and severity analysis through medical imaging.
\newblock In \emph{European Conference on Computer Vision}, page 677–690. Springer, 2022.

\bibitem[Kollias et~al.(2023{\natexlab{a}})Kollias, Arsenos, and Kollias]{kollias2023ai}
Dimitrios Kollias, Anastasios Arsenos, and Stefanos Kollias.
\newblock Ai-enabled analysis of 3-d ct scans for diagnosis of covid-19 \& its severity.
\newblock In \emph{2023 IEEE International Conference on Acoustics, Speech, and Signal Processing Workshops (ICASSPW)}, page 1–5. IEEE, 2023{\natexlab{a}}.

\bibitem[Kollias et~al.(2023{\natexlab{b}})Kollias, Arsenos, and Kollias]{kollias2023deep}
Dimitrios Kollias, Anastasios Arsenos, and Stefanos Kollias.
\newblock A deep neural architecture for harmonizing 3-d input data analysis and decision making in medical imaging.
\newblock \emph{Neurocomputing}, 542:\penalty0 126244, 2023{\natexlab{b}}.

\bibitem[Kollias et~al.(2024)Kollias, Arsenos, and Kollias]{kollias2024domain}
Dimitrios Kollias, Anastasios Arsenos, and Stefanos Kollias.
\newblock Domain adaptation, explainability \& fairness in ai for medical image analysis: Diagnosis of covid-19 based on 3-d chest ct-scans.
\newblock \emph{arXiv preprint arXiv:2403.02192}, 2024.

\bibitem[Subramanian et~al.(2022)Subramanian, Elharrouss, Al-Maadeed, and Chowdhury]{subramanian2022review}
Nandhini Subramanian, Omar Elharrouss, Somaya Al-Maadeed, and Muhammed Chowdhury.
\newblock A review of deep learning-based detection methods for covid-19.
\newblock \emph{Computers in Biology and Medicine}, 143:\penalty0 105233, 2022.

\end{thebibliography}
}

% WARNING: do not forget to delete the supplementary pages from your submission 
% \input{sec/X_suppl}

\end{document}